\RequirePackage{fix-cm}
\documentclass[twocolumn]{svjour3}          
\smartqed  
\usepackage{amssymb}
\usepackage{amsmath, graphicx,amssymb,epsfig,psfrag,subfigure,bm,bbm,booktabs}
\usepackage{color}

\newcommand{\beq}{\begin{equation}}
\newcommand{\eeq}{\end{equation}}
\newcommand{\bpm}{\begin{pmatrix}}
\newcommand{\epm}{\end{pmatrix}}
\newcommand{\beqa}{\begin{eqnarray}}
\newcommand{\eeqa}{\end{eqnarray}}
\newcommand{\beqas}{\begin{eqnarray*}}
\newcommand{\eeqas}{\end{eqnarray*}}

\renewcommand{\d}{\mathrm{d}}

\newcommand{\pdhfrac}[2]{\mathchoice{\frac{#1}{#2}}{#1/#2}{#1/#2}{#1/#2}}

\newcommand{\pd}[2]{\pdhfrac{{\partial}#1}{{\partial}#2}}

\newcommand{\be}{\beta}

\def\XXint#1#2#3{{\setbox0=\hbox{$#1{#2#3}{\int}$ }
\vcenter{\hbox{$#2#3$ }}\kern-.6\wd0}}

\journalname{Structural and Multidisciplinary Optimization}

\begin{document}

\title{On speeding up an asymptotic-analysis-based homogenisation scheme for designing gradient porous structured materials using a zoning strategy}

\titlerunning{Designing graded microstructures with a zoning strategy}        

\author{Dingchuan Xue \and Yichao Zhu* \and Shaoshuai Li \and Chang Liu
\and Weisheng Zhang \and Xu Guo* 
}


\institute{D. C. Xue \and S. S. Li \and C. Liu \and Y. C. Zhu* \and W. S. Zhang \and X. Guo*
          \at State Key Laboratory of Structural Analysis for Industrial Equipment, Department of
           Engineering Mechanics, Dalian University of Technology, Dalian, 116023, P. R. China\\
           International Research Center for Computational Mechanics, Dalian University of Technology \\
           \emph{Present address} of D. C. Xue: Department of Engineering Science and Mechanics, Pennsylvania State University, University Park, Pennsylvania 16802, USA\\
           \email{yichaozhu@dlut.edu.cn}\\
           \email{guoxu@dlut.edu.cn}
}


\date{Received: date / Accepted: date}

\maketitle

\begin{abstract}
  Gradient porous structured materials possess significant potential of being applied in many engineering fields. To accelerate the design process of infill graded microstructures, a novel asymptotic homogenisation topology optimisation method was proposed by Zhu et al. \cite{ZhuYC_JMPS2019}, aiming for 1) significantly enriching the pool of representable graded microstructures; 2) deriving an homogenised formulation for stress analysis in consistency with fine-scale results. But the work is severely confined from being widely applied, mainly due to the following two reasons. Firstly, to circumvent the macroscopically pointwise computation for solving various microscopic cell problems, linearisation had to be adopted for its numerical implementation, and this significantly reduces the design freedom. Secondly, lacking of sensitive analysis, genetic algorithm was chosen for optimisation, inevitably decreasing the computational efficiency. To address these bottleneck challenging issues, a zoning scheme empowered by computational parallelism is introduced, and the sensitivity analysis associated with the new asymptotic framework is conducted. Through comparisons with fine-scale simulation results, the proposed algorithm is shown to be an effective tool for evaluating the mechanical behaviour of graded microstructures. As an optimisation tool, the mapping function takes a concise and explicit form. But its parameterisation still needs further investigation, so as to improve the solution optimality of the present approach, especially in comparison with another recently proposed method \cite{Groen_IJNME2018}.
  The high computational efficiency shown by the proposed scheme is also demonstrated with optimisation results for three-dimensional graded microstructures, which are not frequently discussed in literature, possibly because of the high computational cost generated.

\keywords{Graded microstructures \and Topology optimisation \and Asymptotic analysis \and Homogenisation \and Moving morphable components}
\end{abstract}

\section{Introduction}
Functional equipments and devices infilled with well-designed microstructures have started to see their applications in mechanical, thermal, acoustics and optical aspects \cite{Lakes_Nature1993,DongHW_JMPS2017,Sigmund_APL1996,Kushwaha_PRL1993,LiuC_PRAppl2015,Vogiatzis_CMAME2018}. Driven by the development of additive manufacturing techniques, the fabrication of complex microstructures has become increasingly accessible, and this naturally stimulates the development of the corresponding intelligent/automatic design algorithms. Among them, topology optimisation (TO) methods, which aim at optimally distributing a certain amount of materials, within a design domain, for achieving optimised properties/functions, have demonstrated their effectiveness for microstructural design \cite{Aage_Nature2017} and have arose considerable research interest in recent years.

Computational approaches for microstructural design normally fall into the category of multiscale topology optimisation, i.e., at least two length scales are involved: a macroscopic scale on which the overall structural response is concerned, and a microscopic scale on which the structural members are resolved. Doing topology optimisation straightforward on a fine-scale level, albeit a number of valuable works conducted \cite{Alexandersen_CMAME2015,LiuC_JAM2017}, usually poses huge demands on computational resources. One way to manage this efficiency-versus-accuracy dilemma, is to develop multiscale approaches in a viewpoint of scale separation. The basic idea lying behind the scale-separation treatments is to envisage the design domain as a macroscopic continuum, and the locally homogenised properties of the continuum are quantified by solving a set of boundary value problems defined within a microscopic cell of rectangle/cuboid shape.

The mathematical foundation beneath the the aforementioned scale-separation methods, is the asymptotic homogenisation (AH) results for periodic structures \cite{Bensoussan1978}. The idea of AH was pioneerly implemented for structural design by Bendsoe and Kichuchi \cite{Bendsoe_CMAME1988}, where each macroscopic pixel defines a microscopic cell. Restricting the volume fraction at each pixel to take binary values (0 or 1) by means of penalisation introduced the nowadays widely used solid isotropic materials penalty (SIMP) framework \cite{Bendsoe_StructOpt1989,ZhouM_CMAME1991,Sigmund_IJSS1994,Aage_Nature2017}. The treatment was then developed for structural design taking place on concurrently multiple scales \cite{Rodrigues_SMO2002,Coelho_SMO2008,LiuL_CompStruct2008}, and was further generalised for multi-functional designs \cite{NiuB_SMO2009,DengJD_SMO2013,YanJ_CompMech2016}, as well as for the consideration of loading uncertainty \cite{DengJD_SMO2017}, etc..

The aforementioned AH-based works provide an (asymptotically) accurate and coarse-grained formulation for analysing the mechanical behaviour of configurations filled with spatially periodic microstructures. However, a considerably huge amount of high-performance microstructures, both naturally formed \cite{Sanchez_NatMater2005,Fratzl_Nature2009,Meyers_Sci2013} and manufactured \cite{Jorgensen_IJSS1998,Arabnejad_JBiomedEng2012,Cheng_Sci2018}, displays a spatially varying nature. For optimal design of such configurations filled with graded microstructures or ``graded microstructural configurations''(GMCs), further studies are needed in the following three aspects: GMC representation, its stress analysis, as well as its topology optimisation. For GMC representation, a set of interesting works have been recently carried out by letting the topology description functions (TDFs) taking different level-set values \cite{WangYQ_CMAME2017,ZhangY_CompMaterSci2018,ZhangY_SMO2019}. As for the stress analysis and topology optimisation of GMC, considerable efforts have been made through modification of the existing AH formulations, which are originally devised merely for spatially periodic configurations \cite{ZhouSW_JMaterSci2008,Radman_JMaterSci2013,Radman_CompMaterSci2014,ChengL_CMAME2019,Gao_AES2018}. Such treatments generally see their limitations in three aspects: 1) constituting cells are normally of rectangle or cuboid shape, thus the spatial variance is only permitted along directions parallel to one of the cell edges; 2) smooth connectivity across cell boundaries can not be ensured; 3) the computational accuracy of the calculated structural responses can not be controlled.

To circumvent these challenging issues in the use of AHTO for optimising GMC compliance, a de-optimisation topology optimisation scheme was recently proposed \cite{Groen_IJNME2018,Groen_CMAME2019,WuJ_IEEE2019,Allaire_CompMathAppl2019,GeoffroyDonders_JCP2020,Groen_arXiv2019}. They first search for an optimal continuum distribution of laminar orientations, whose homogenised elastic properties are obtained either by AH or microscale pre-computation. Then the GMC's details are fully resolved with the use of a so-called projection method \cite{Pantz_SIAMJCO2008}, that is, by numerically projecting the laminates in alignment with the desired local principal stress. Such de-homogenisation approaches witness breakthrough in overcoming the above-mentioned issues limiting traditional asymptotic-homogenisation topology optimisation (AHTO) methods. Along this direction, two-dimensional (2D) cases have been systematically analysed \cite{Groen_IJNME2018,Groen_CMAME2019,Allaire_CompMathAppl2019}, and attentions to three-dimensional (3D) cases have been paid fairly recently \cite{WuJ_IEEE2019,GeoffroyDonders_JCP2020,Groen_arXiv2019}.

An alternative route to substantially improve the AHTO methods for GMC design, is to systematically re-derive the asymptotic homogenisation results suitable for analysing the mechanical behaviour of GMC. This motivates the work by Zhu et al. \cite{ZhuYC_JMPS2019}. They performed asymptotic analysis to re-build the mathematical foundation of the original framework, so that the description, compliance computation and topology optimisation of GMC are harmonically integrated in the developed framework. In order to distinguish it from the traditional AHTO formulation, the solution framework \cite{ZhuYC_JMPS2019} is termed as ``AHTO plus" in this article. Within the AHTO plus framework. the spatial variance of GMC is achieved by using a function composition of a macroscopic mapping function with a locally periodic TDF for the microscopic cell configuration. Besides, the constituting cells in the AHTO plus framework are not necessarily confined to take either rectangle / cuboid shape or laminate configurations.

In theory, the method of AHTO plus established a solid theoretical foundation to address the bottleneck issues limiting the conventional AHTO approaches for effective GMC optimisation \cite{ZhuYC_JMPS2019}. From a practical viewpoint, however, improvement is still highly desired, mainly due to the following two reasons. Firstly, no explicit formulation of sensitivity analysis has been derived in accordance with the AHTO plus approach. Secondly, to maintain high accuracy, one has to compute the microscopic equilibrium problem as many times as the number of the finite elements (FEs) used for computing the macroscopic structural response in each iteration step. Due to the aforementioned two aspects of limitation, Zhu et al. \cite{ZhuYC_JMPS2019} had to adopt a less efficient genetic algorithm and only tested the linearised form of the AHTO plus formulation for numerical implementation.

In face of these challenging issues lying on the application side of the AHTO plus formulation, this article is presented. Note that the most time-consuming part when implementing the AHTO plus method, is to solve the microscopic cell problems repeatedly. But these cell problems are actually defined in a macroscopically pointwise sense. To make the best use of this feature, a zoning strategy is proposed. By decomposing the macroscopic computational domain into several zones, the evaluation of the microscopic cell problems can be conducted in a fully parallel way. This significantly improves the computational efficiency. In fact, detailed analysis will be carried out to thoroughly assess the efficiency brought by the use of the proposed zoning strategy, and this is the first objective of the present work. The second issue here is to perform sensitivity analysis associated with the AHTO plus formulation, which further speeds up the proposed algorithm. To demonstrate the proposed method as an effective tool for analysing the mechanical behaviour of GMC, coordinated simulation results regarding a two-dimensional case are present. Then the performance of AHTO plus formulation, as a tool for the compliance optimisation of GMC is examined, through a comparison with the benchmark results obtained by using the de-homogenisation method \cite{Groen_CMAME2019}. It is found that the compliance value computed with the present method is still 20\% higher than its counterpart by \cite{Groen_CMAME2019}, and key issues for improvement are then discussed. Finally, we will show simulation results of three-dimensional GMCs, which are not frequently been examined in literature, possibly because of the high computational cost. We find that, with the use of the present zoning scheme, it should take an eight-processor computer few hours to output an optimised design.

The article is arranged as follows. In Sec.~\ref{Sec_AHTO_review}, the AHTO plus formulation is firstly outlined, and the challenging issues limiting its potential application are discussed. Then in Sec.~\ref{Sec_introduction_zoning}, we introduce the zoning strategy empowered by parallel computation, which aims at resolving the challenging issues associated with the original AHTO plus approach. The gradient-based sensitivity results are also derived in Sec.~\ref{Sec_introduction_zoning}. In Sec.~\ref{Sec_numerical_examples}, both 2D and 3D examples are presented, and the key features of the proposed scheme are examined in depth. The article concludes with a discussion session in Sec.~\ref{Sec_conclusion}.

In this article, indexation is needed in various occasions with the following rules applied. English letters in lower cases are used for the indexation associated with spatial dimensions; English letters in upper cases are used for the indexation associated with subdomains; Greek letters are used for the indexation associated with design variables. In this work, the Einstein summation rule is applied only for indexation of English letters.

\section{Outline of AHTO plus formulation and challenging issues associated with its numerical implementation\label{Sec_AHTO_review}}
In this section, we shall review the AHTO plus formulation \cite{ZhuYC_JMPS2019}, which comprises of three parts: the GMC representation, the homogenisation formulae calculating its mechanical behaviour, and the associated topology optimisation formulation. Then the challenging issues limiting the efficient numerical implementation of the AHTO plus approach are addressed.

\subsection{Outline of the AHTO plus formulation}
\subsubsection{Multiscale representation of a configuration filled with quasi-periodic graded microstructures}
\begin{figure*}[!ht]
  \centering
  \includegraphics[width=.6\textwidth]{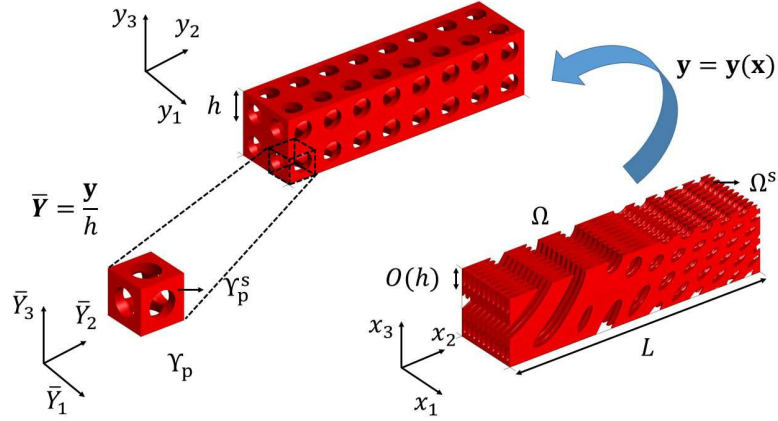}
  \caption{A configuration filled with graded microstructures and the treatment of mapping it to a periodic configuration in a fictitious space measured in $\mathbf{y}$.}
  \label{GMs}
\end{figure*}

Graded microstructural configurations normally involve multiple length scales. For a configuration filled with graded microstructures, with one example shown in the bottom-right panel of Fig.~\ref{GMs}, its constituting members are characterised by a length-scale parameter $h$, while it has also an overall size characterised by another length-scale parameter $L$. In general,
\beq\label{scale_separation}
\frac{h}{L} \triangleq \epsilon \ll 1.
\eeq
In this article, we term the length scale characterised by $h$ the ``microscale'', and term the length scale characterised by $L$ the ``macroscale''.

Within any topology optimisation framework, a GMC in $\mathbb{R}^n$ can be described by a topology description function $\chi(\mathbf{x})$. As shown in the bottom-right panel of Fig.~\ref{GMs}, we denote the overall porous region by $\Omega$, and denote the solid part in $\Omega$ by $\Omega^\text{s}$.
Then a TDF can be assigned in the whole design domain $\Omega$, such that $\chi(\mathbf{x})\geq0$ for all $\mathbf{x}\in\Omega^\text{s}$ and $\chi(\mathbf{x})<0$ for all $\mathbf{x}\in\Omega\setminus\Omega^\text{s}$.

Given the normally large contrast in the two length scales, extremely fine mesh size (far shorter than $h$) is necessary in order to resolve a GMC. This requires a relatively large data set to store it. To overcome this issue, a multiscale scheme for representing quasi-periodic graded microstructures was introduced \cite{LiuC_JAM2017} and further modified \cite{ZhuYC_JMPS2019}. The key idea is illustrated in Fig.~\ref{GMs}. For a GMC as shown in the bottom-right panel of Fig.~\ref{GMs}, a macroscopically smooth mapping function $\mathbf{y}=\mathbf{y}(\mathbf{x})$ is defined, so as to map it to a spatially periodic configuration of uniform period $h$ (as shown in the middle panel of Fig.~\ref{GMs}). For expressing this periodic configuration defined in the fictitious space measured in $\mathbf{y}$, one simply assigns a periodic TDF, say $\chi^\text{p}(\frac{\mathbf{y}}{h})$, to one of its representative cells, as shown in the bottom-left panel of Fig.~\ref{GMs}, and then replicates it in the fictitious space. Here the TDF $\chi^\text{p}(\frac{\mathbf{y}}{h})$ effectively resolves the microstructural details (in the fictitious space), and the microscale parameter $h$ appearing in the denominator, is to rescale its domain of definition to a non-dimensional unit cell denoted by $\Upsilon_\text{p}=[0,1]^n$ ($n$=2 or 3 unless otherwise stated). Finally, the TDF for the GMC in the actual space measured by $\mathbf{x}$ is represented by the composition of the macroscopically smooth mapping function $\mathbf{y}(\mathbf{x})$ and the microscale TDF $\chi^\text{p}(\bullet)$ defined in the representative unit cell, i.e.,
\beq\label{Map}
\chi(\mathbf{x})=\chi^\text{p}\left(\frac{\mathbf{y}(\mathbf{x})}{h}\right).
\eeq
Given $\chi^\text{p}(\bullet)$'s periodicity for its entry, the TDF defined by Eq.~\eqref{Map} automatically ensures the smooth connection within GMCs.

Although in a seemingly simple form, the multiscale representation given by Eq.~\eqref{Map} has seen its abundance in its describable GMCs, and one may refer to \cite{ZhuYC_JMPS2019} for more illustrative examples. Besides, Eq.~\eqref{Map} enables one to squeeze the digital space needed for memorising a GMC. This is because fine mesh is only necessary for representing the TDF $\chi^\text{p}(\bullet)$ in one unit cell, while the mapping function $\mathbf{y}(\mathbf x )$, given its global smoothness, can be stored on a grid far coarser than the microscale parameter $h$.

\subsubsection{Asymptotic analysis of elasticity problems involving GMC\label{Sec_AA}}
The governing equation of a typical elasticity problem involving GMC can be formulated by
\beq\label{equilibrium_non_homo}
\pd{}{x_j}\left(\mathbb{C}_{ijkl}\pd{u_k}{x_l}\right)+f_i=0,\, \text{in }\Omega^\text{s}
\eeq
for $i=1$, $\cdots$, $n$, where $\mathbf{u}=(u_i)\in\mathbb{R}^n$ denotes vector of the displacement field; $\mathbf{f}=(f_i)\in\mathbb{R}^n$ is vector of the body force density and $\mathbb{C}=(\mathbb{C}_{ijkl})~(i,j,k,l=1,\dots,n)$ is the fourth-order elasticity tensor of the solid materials constituting the GMC. Given its multiscale nature, evaluating the stress response of a GMC usually requires extremely fine FE meshes, and this is often computationally intractable.

To address this issue, the continuum limit of Eq.~\eqref{equilibrium_non_homo}, as the ratio $\epsilon$ tends to zero, was studied \cite{ZhuYC_JMPS2019}. In the presence of microstructures, the spatial variance in the displacement field $\mathbf{u}$ should take place concurrently on two scales. One is local oscillation accommodating the microstructural change on the microscale, and the other is a global trend accommodating the external load on the macroscale. To capture the locally oscillatory feature, a non-dimensional spatial variable
\begin{equation} \label{Y_def}
\bar{\mathbf{Y}} = \frac{\mathbf{y}(\mathbf{x})}{h},
\end{equation}
is introduced with $\bar{\mathbf{Y}}\in\Upsilon_{\text{p}}=[0,1]^n$. Then the displacement field is approximated by a series expanded in terms of $\epsilon$:
\begin{equation} \label{u_expansion}
\mathbf{u} \sim \mathbf{u}^{(0)}(\mathbf{x};\bar{\mathbf{Y}}) + \epsilon \mathbf{u}^{(1)}(\mathbf{x};\bar{\mathbf{Y}}) + \cdots,
\end{equation}
where the spatial variable $\mathbf{x}$ is maintained only to measure smooth variance in space. Since the GMC becomes periodic in the fictitious space of $\mathbf{y}$, one may assume that $\mathbf{u}^{(i)}$, $i=0$, $1$, $2$, $\cdots$, are all periodic in $\bar{\mathbf{Y}}\in[0,1]^n$. Thus the originally multiscale displacement field is now asymptotically reproduced by a series with the two scales separated. The present treatment is generalised from the traditional homogenisation works for periodic configurations (in actual space) \cite{Bensoussan1978,Cioranescu2010}, where no mapping function is necessary.

Then the asymptotic behaviour of the system is investigated until the equation for $\mathbf{u}^{(2)}$ \cite{ZhuYC_JMPS2019}. They firstly showed that $\mathbf{u}^{(0)}$ is actually independent of $\bar{\mathbf{Y}}$, and it can be treated as the ``homogenised'' displacement field denoted by $\mathbf{u}^{\mathrm{H}}$. Here a superscript ``$\mathrm{H}$'' is affiliated with an variable to indicate that it is a homogenised quantity (independent of $\bar{\mathbf{Y}}$).

Thus the original multiscale problem is (asymptotically) re-formulated by a homogenised problem where the region $\Omega$ is envisaged as a solid continuum. This effectively gives rise to the macroscopic force-balance equation:
\beq\label{equilibrium_homo}
\pd{}{x_j}\left(\mathbb{C}^\mathrm{H}_{ijkl}\pd{u_k^\mathrm{H}}{x_l}\right)+f_i^\mathrm{H} =0,\, \mathbf{x}\in \Omega
\eeq
for $i=1$, $\cdots$, $n$ with appropriate homogenised boundary conditions imposed, where $\mathbb{C}^\mathrm{H}=\left(\mathbb{C}^\mathrm{H}_{ijkl}\right)$ ($i$, $j$, $k$, $l=1$, $\cdots$, $n$) is the homogenised elasticity tensor of the equivalent continuum; $\mathbf{f}^\mathrm{H} = \left(f^\mathrm{H}_i\right)$ ($i=1$, $\cdots$, $n$) measures the homogenised body force density. In the present work, we set $\mathbf{f}^\mathrm{H} = \boldsymbol{0}$. The homogenised elasticity tensor $\mathbb{C}^\mathrm{H}$ is calculated by
\beq\label{homogenised elasticity}
\mathbb{C}_{ijkl}^\mathrm{H}=\mathbb{C}_{ijkl}\left|\Upsilon^\text{s}_\text{p}\right| - \mathbb{C}_{ijst}J_{nt}\int_{\Upsilon^\text{s}_\text{p}}\pd{\xi^{kl}_s}{\bar Y_n}\d \bar{\mathbf{Y}},
\eeq
where $\Upsilon^\text{s}_\text{p}$ denotes the (non-dimensional) porous solid region within the unit generating cell $\Upsilon_\text{p}$, and $\left|\Upsilon_\text p ^\text s\right|$ denotes its measure; $\mathbf J=\left(J_{ij}\right)$ ($i$, $j=1$, $\cdots$, $n$) is the Jacobean matrix associated with the mapping function
\beq\label{Jacobean matrix}
J_{ij}=\pd{y_i}{x_j}
\eeq
measuring the degree of spatial variance of the GMC; $\boldsymbol{\xi}=\left(\xi^{ij}_k\right)$ $(i,j,k=1,\dots,n)$ is a third-order tensor. It satisfies a set of microscale cell problems given by
\beq\label{equilibrium_micro}
J_{mj}\pd{}{\bar Y_m}\left(\tilde{\mathbb{C}}_{ijkl}J_{nl}\pd{\xi_k^{st}}{\bar Y_n}\right)= J_{mj}\pd{\tilde{\mathbb{C}}_{ijst}}{\bar Y_m},\quad \bar{\mathbf{Y}}\in \Upsilon_\text{p},
\eeq
for $s$, $t=1$, $\cdots$, $n$, with periodic boundary conditions imposed, where $\tilde{\mathbb{C}}_{ijkl}$ is given by
\beq \label{C_bar}
\tilde{\mathbb{C}}_{ijkl} = \left\{\begin{aligned}
& \mathbb{C}_{ijkl}, \quad && \bar{\mathbf{Y}}\in \Upsilon_\text{p}^\text{s};\\
& 0, && \bar{\mathbf{Y}} \in \Upsilon_\text{p}\setminus \Upsilon_\text{p}^\text{s}.
\end{aligned}\right.
\eeq

Once the homogenised displacement field $\mathbf{u}^\mathrm{H}$ is obtained, the homogenised stress field can be computed by
\beq\label{Constitutive equation}
\sigma_{ij}^\mathrm{H}=\mathbb{C}_{ijkl}^\mathrm{H}u_{k,l}^\mathrm{H},
\eeq
for $i$, $j=1$,$\cdots$, $n$. Then the homogenised structural compliance is given by
\begin{equation} \label{homogenised_compliance}
\mathcal{C}^\mathrm{H}=\int_\Omega \mathbb{C}_{ijkl}^\mathrm{H}\pd{u_i^\mathrm{H}}{x_j}\pd{u_k^\mathrm{H}}{x_l}\,\d \mathbf{x}.
\end{equation}

In theory, as $\epsilon\rightarrow0$, $\sigma_{ij}^\mathrm{H}$ should be weakly convergent to the corresponding stress component calculated from the original multiscale problem~\eqref{equilibrium_non_homo}, and $\mathcal{C}^\mathrm{H}$ should converge to the actual GMC compliance. This conclusion, without rigorous proofs so far, is drawn in analogy with the proof for spatially periodic cases \cite{Cioranescu2010}.

\subsubsection{The optimisation formulation}
An optimisation framework is then established through minimising the homogenised structural compliance as follows
\beq \label{homogenised optimisation}
\begin{aligned}
&\text{Find}\qquad \mathbf y=\mathbf y(\mathbf x)\in\mathcal U_\mathbf y, \,\chi(\bar{\mathbf{Y}}) \in \mathcal{U}\left(\Upsilon_{\text{p}}\right). \\
&\text{Minimise}\qquad \mathcal{C}^\mathrm{H}=\int_\Omega \mathbb{C}_{ijkl}^\mathrm{H}\pd{u_i^\mathrm{H}}{x_j}\pd{u_k^\mathrm{H}}{x_l}\,\d \mathbf{x},\\
&\text{s.t.}\\
&\int_\Omega \mathbb{C}_{ijkl}^\mathrm{H}u_{i,j}v_{k,l}\d \mathbf{x} = \int_\Omega f_i^\mathrm{H}v_i \d \mathbf{x} + \int_{\Gamma_{\text{t}}} t_iv_i\d S,\\
&\qquad\qquad\qquad\qquad\qquad\qquad\qquad\qquad\qquad\forall\,\mathbf v \in \mathcal U_{ad};\\
&\mathbb{C}_{ijkl}^\mathrm{H}=\mathbb{C}_{ijkl}\left|\Upsilon^\text{s}_\text{p}\right| -\mathbb{C}_{ijst}J_{nt}\int_{\Upsilon^\text{s}_\text{p}}\pd{\xi^{kl}_s}{\bar Y_n}\d \bar{\mathbf{Y}};
\\
&J_{mj}\pd{}{\bar Y_m}\left(\tilde{\mathbb{C}}_{ijkl}J_{nl}\pd{\xi_k^{st}}{\bar Y_n}\right)= J_{mj}\pd{\tilde{\mathbb{C}}_{ijst}}{\bar Y_m},\, \bar{\mathbf{Y}}\in \Upsilon_\text{p};
\\
&\mathbf u^{\text{H}} = \bar {\mathbf u},\quad \text{on }\Gamma_{\text{u}};\\
& \text{vol.} \le \bar{V},
\end{aligned}
\eeq
where $\mathbf v$ is a virtual displacement field; $\mathcal{U}_{\mathbf{y}}$ and $\mathcal{U}(\Upsilon_{\text{p}})$ are the function spaces for the macroscopic mapping functions and the microscale TDF, respectively; $\Gamma_{\text{u}}$ is the boundary part of $\Omega$ on which displacement boundary condition is imposed; $\Gamma_{\text{t}}$ is the boundary part of $\Omega$ on which a traction of $\mathbf{t}$ is applied, and $\mathbf{n}$ is outer normal vector on the domain boundary $\partial \Omega$; ``vol.'' denotes the volume fraction of the solid region; $\bar{v}$ is the upper bound of this volume fraction.

Compared to traditional asymptotic-homogenisation topology optimisation methods \cite{Bendsoe_CMAME1988,Bendsoe_StructOpt1989,Rodrigues_SMO2002,Coelho_SMO2008,NiuB_SMO2009}, the formulation presented here in \S~\ref{Sec_AA} ensures smooth spatial variance in GMC by gradually deforming the matrix structure $\Upsilon_{\text{p}}^{\text{s}}$ with regards to the mapping function $\mathbf{y} = \mathbf{y}(\mathbf{x})$. It is thus termed as an ``AHTO plus'' method. Compared to the projection-based methods \cite{Pantz_SIAMJCO2008,Groen_IJNME2018,Groen_CMAME2019,Allaire_CompMathAppl2019,GeoffroyDonders_JCP2020,Groen_arXiv2019}, the framework outlined by Eq.~\eqref{homogenised optimisation} employs a different way to enable spatial variance. In the projection-based approaches, the shape and rotating angle of the unit cell $\Upsilon_{\text{p}}$ vary in space. In the AHTO plus framework, the unit cell stays uniform in space, but it can rotate or deform itself in space to ensure GMC's varieties. Further discussion over the comparison between the different methods will be carried out in Sec.~\ref{Sec_comparison}.

\subsection{Challenges in the implementation of the AHTO plus approach}
At its present stage, the AHTO plus approach \cite{ZhuYC_JMPS2019} still struggles for being effectively applied, mainly for two reasons.

Firstly, the AHTO plus formulation involves solving two linear partial differential equation systems: the macroscale equilibrium equation \eqref{equilibrium_homo} for the homogenised displacement field $\mathbf{u}^\mathrm{H}$, and the microscale equilibrium equations \eqref{equilibrium_micro} for the third-order tensor $\boldsymbol{\xi}=\left(\xi^{ij}_k\right)$, $i$, $j$ and $k=1$, $\cdots$, $n$. Note that Eq.~\eqref{equilibrium_micro} includes $\mathbf J=\left(J_{ij}\right)$, the Jacobean matrix measuring the degree of GMC's spatial variance, which is a function of the macroscopic position $\mathbf x$. Thus one needs to solve the microscopic equilibrium equation \eqref{equilibrium_micro}, as many times as the number of finite elements used for solving the macroscopic equilibrium equation \eqref{equilibrium_homo}. This severely reduces the overall computational efficiency. To circumvent this challenging issue, only the linearised form of the AHTO plus formulation was considered for simulation \cite{ZhuYC_JMPS2019}. Upon linearisation, the computed compliance is only accurate for GMCs that are produced by slightly perturbing spatially-periodic configurations. As a result, the design freedom shrinks significantly. For example, the linearised AHTO plus method simply outputs an optimised two-dimensional cantilever filled with graded microstructures, with a compliance value three times larger than the result computed by using the de-homogenisation method \cite{Groen_IJNME2018}.

Secondly, no sensitivity analysis was considered with respect to the AHTO plus formulation thus far. As a result, one has to employ resource-demanding algorithms, such as genetic algorithm, for optimisation \cite{ZhuYC_JMPS2019}. This also inevitably affects efficient numerical implementation of the AHTO plus method.

Facing these challenges, this article is presented. In consideration of the AHTO plus method's generic favour for parallel computation, a zoning strategy is proposed to accelerate the time-consuming computation of the cell problems defined by Eq.~\eqref{equilibrium_micro}. Moreover, sensitivity analysis is also carried out, so as to enable more efficient gradient-based algorithms.

\section{A zoning strategy and sensitivity analysis\label{Sec_introduction_zoning}}
In this section, a zoning scheme empowered by computational parallelism is proposed for speeding up the numerical implementation of the AHTO plus formulation. Then sensitivity results within the AHTO plus framework are are derived. The present section concludes with a flow chart illustrating the present algorithm.

\subsection{Domain decomposition with parallel computing\label{Sec_theory_parallel}}
The most time-consuming part when implementing the AHTO plus formulation, as mentioned above, is to solve the (macroscopically) pointwise microscale cell problems governed by Eq.~\eqref{equilibrium_micro}. The reason is that Eq.~\eqref{equilibrium_micro} depends on $\mathbf J(\mathbf x)$, which varies within the homogenised continuum region $\Omega$. Hence one has to solve Eq.~\eqref{equilibrium_micro} with respect to each finite element.

\begin{figure}[!ht]
  \centering
  \includegraphics[width=.48\textwidth]{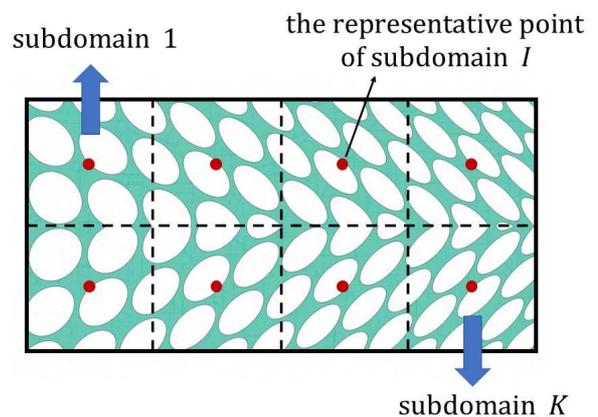}
  \caption{Domain decomposition.}
  \label{Zoning}
\end{figure}

The proposed zoning strategy is to divide the macroscopic region $\Omega$ into $K$ subdomains as shown in Fig.~\ref{Zoning}.
Then the macroscopic continuum is considered as constituting $K$ subdomains made of homogenous materials. The effective elasticity tensor associated with each subdomain, denoted by $\mathbb{C}^I=\mathbb{C}_{ijkl}^I$, takes constant values. To derive $\mathbb{C}_{ijkl}^I$, a representative point associated with each subdomain, say, $\mathbf x^I$ for the subdomain $\Omega^I$, is selected as shown in Fig.~\ref{Zoning}. Then we use the homogenised elasticity tensor evaluated at $\mathbf x^I$ to represent the elastic properties of the whole subdomain. Mathematically, this is done by defining $\mathbf J^I=\mathbf J(\mathbf x^I)$. Then the cell problem for the third-order tensor $\boldsymbol{\xi}=\left(\xi_k^{st,I}\right)$, $k$, $s$, $t=1$, $\cdots$, $n$, is written by
\beq\label{equilibrium_micro_zoning}
J_{mj}^I\pd{}{\bar Y_m}\left(\tilde{\mathbb{C}}_{ijkl}J_{nl}^I\pd{\xi_k^{st,I}}{\bar Y_n}\right)= J_{mj}^I\pd{\tilde{\mathbb{C}}_{ijst}}{\bar Y_m},\, \bar{\mathbf{Y}}\in [0,1]^n
\eeq
with periodic boundary conditions imposed, where $i=1,\dots,n,$ and $I=1,\dots,K$. Note that the Einstein summation rule is not valid for the index $I$ for identifying subdomains.

Therefore, the homogenised elasticity tensor becomes a tensor of piecewisely constant values. In the $I$-th subdomain, the corresponding term, denoted by $\mathbb{C}_{ijkl}^{I}$ is calculated by
\beq\label{homogenised elasticity_zoning}
\mathbb{C}_{ijkl}^{I} = \mathbb{C}_{ijkl}\left|\Upsilon^\text{s}_\text{p}\right| -\mathbb{C}_{ijst}J_{nt}^I\int_{\Upsilon_\text{p}}\pd{\xi^{kl,I}_s}{\bar Y_n}\d \bar{\mathbf{Y}}.
\eeq
Compared with the original cell problem governed by Eq.~\eqref{equilibrium_micro}, the coefficients in Eq.~\eqref{equilibrium_micro_zoning} which are contained in $\mathbf J^I$ no longer depend on the macroscopic position. This means that one simply needs to solve Eq.~\eqref{equilibrium_micro} $K$ times, where $K$ is recalled to be the total number of subdomains.

Then with the homogenised displacement field $\mathbf{u}^{\text{H}}$ (approximately) computed from the macroscopic equilibrium equation \eqref{equilibrium_homo}, the homogenised GMC compliance of GMC is approximated by
\beq\label{Compliance_zoning}
\mathcal{C}^\mathrm{H} \approx \sum_{I=1}^K\int_{\Omega^I} \mathbb{C}_{ijkl}^{I}\pd{u_i^\mathrm{H}}{x_j}\pd{u_k^\mathrm{H}}{x_l}\d \mathbf{x}.
\eeq

In theory, the compliance computed from this zoning scheme should converge to that obtained from the original formulation \eqref{homogenised optimisation}, as the subdomains become more coincident with the FE mesh for computing the macroscopic problem. Practically, a good convergence is seen in a two-dimensional example (to be presented in Sec.~\ref{Sec_result_analysis}, as one subdomain covers at least 25 FEs).

When the microscopic problem~\eqref{equilibrium_micro_zoning} is solved, there is no data exchange between computational processes carried out in different subdomains. This indicates that the zoning scheme may be more efficient, if parallel computing strategy is employed. If we define a single ``task'' by the process of calculating $\mathbb{C}^I_{ijkl}$ in one subdomain, then one can assign on each cluster equal number of tasks, and run them in parallel. Moreover, for each task, the input is $\mathbf J^I$ and the output is $\mathbb{C}_{ijkl}^{I}$. They are both independent of the detailed mesh within the unit cell $\Upsilon_{\text{p}}$. So when $\mathbb{C}_{ijkl}^{I}$ is obtained, there is no need to keep the values for the third-order tensor $\xi_k^{ij,I}$. This helps save memories in simulation. It will be numerically shown in Sec.~\ref{Sec_result_analysis} that the simulations get accelerated dramatically with the use of the proposed zoning scheme.

\subsection{Gradient-based sensitivity analysis\label{Sec_sensitivity}}
To further speed up the optimisation process, the gradient-based sensitivity analysis in accordance with the AHTO plus formulation is also carried out. Here the sensitivity of the homogenised compliance is expressed by
\beq\label{sensitivity_C}
\pd{\mathcal{C}^\mathrm{H}}{d_\gamma}=-\sum^K_{I=1}\int_{\Omega}\pd{ \mathbb{C}^I_{ijkl}}{d_\gamma}\pd{u^\mathrm{H}_i}{x_j}\pd{
u^\mathrm{H}_k}{x_l}\d \mathbf{x},
\eeq
where the vector $\mathbf d=(d_1,\dots,d_\lambda)^\top$ contains all the design variables.

The design variables associated with the AHTO plus approach can be classified into two groups: the macroscopic design variables identified by subindices $\alpha$, which parameterise the mapping function $\mathbf{y}(\mathbf{x})$; and the microscopic design variables identified by subindices $\beta$, which are the parameters controlling the TDF $\chi^{\text{p}}(\bullet)$ defined in the representative unit cell.

The sensitivity analysis with respect to the macroscale variables $d_\alpha$ is firstly considered.
For this purpose, we can follow \cite{ZhuYC_JMPS2019} to rewrite $\mathbb{C}^I_{ijkl}$ given by Eq.~\eqref{homogenised elasticity_zoning} by
\beq\label{C_tensor}
\begin{aligned}
\mathbb{C}^I_{ijkl} & =\int_{\Upsilon_\text{p}}\tilde{\mathbb{C}}_{pqst} \left(\delta_{ip}\delta_{jq}-J_{nq}^I\pd{\xi^{ij,I}_p}{\bar Y_n}\right) \times\\
& \qquad\quad \left(\delta_{ks}\delta_{lt}-J_{mt}^I\pd{\xi^{kl,I}_s}{\bar Y_m}\right)\d {\bar{\mathbf{Y}}}.
\end{aligned}
\eeq
Taking the derivatives with respect to $d_\alpha$, $\alpha=1$, $\cdots$, $\lambda_\alpha$ (where $\lambda_\alpha$ is the total number of macroscale design variables), on both sides of Eq.~\eqref{C_tensor} gives
\beq \label{sensitivity_p1}
\pd{\mathbb{C}^I_{ijkl}}{d_\alpha}=\pd{\mathbb{C}^{I[1]}_{ijkl}}{d_\alpha} + \pd{ \mathbb{C}^{I[2]}_{ijkl}}{d_\alpha}
\eeq
where
\beq \label{sensitivity_p1a}
\pd{\mathbb{C}^{I[1]}_{ijkl}}{d_\alpha} = -\int_{\Upsilon_\text{p}} \left[\pd{J_{nq}^I}{d_\alpha}\pd{\xi^{ij,I}_p}{\bar Y_n}\tilde{\mathbb{C}}_{pqkl} + \pd{J_{mt}^I}{d_\alpha}\pd{\xi^{kl,I}_s}{\bar Y_m}\tilde{\mathbb{C}}_{ijst} \right]\d {\bar{\mathbf{Y}}},
\eeq
and
\beq \label{sensitivity_p1b}
\begin{aligned}
\pd{\mathbb{C}^{I[2]}_{ijkl}}{d_\alpha} &= \int_{\Upsilon_\text{p}} \left[\pd{J_{nq}^I}{d_\alpha}\pd{\xi^{ij,I}_p}{\bar Y_n}\left(\tilde{\mathbb{C}}_{pqst}J_{mt}^I\pd{\xi^{kl,I}_s}{\bar Y_m}\right)\right] \, \d {\bar{\mathbf{Y}}} \\
& \quad + \int_{\Upsilon_\text{p}} \left[\pd{J_{mt}^I}{d_\alpha}\pd{\xi^{kl,I}_s}{\bar Y_m}\left(\tilde{\mathbb{C}}_{pqst}J_{nq}^I\pd{\xi^{ij,I}_p}{\bar Y_n}\right) \right]\,\d {\bar{\mathbf{Y}}},
\end{aligned}
\eeq
respectively. Thus one simply needs to evaluate the derivatives of the Jacobean matrix with respect to the macroscale design variables at all representative points $\mathbf{x}^I$. If the mapping function $\mathbf{y}(\mathbf{x})$ is explicitly parameterised by $d_{\alpha}$, then the associated derivatives should be obtained in a straightforward way.

As for the sensitivity with respect to the microscale variables, taking the derivatives with respect to $d_\beta$, $\beta=1$, $\cdots$, $\lambda_\beta$ (where $\lambda_\beta$ is the total number of microscale design variables), on both sides of Eq.~\eqref{homogenised elasticity_zoning} gives
\beq \label{sensitivity_p2}
\pd{\mathbb{C}^I_{ijkl}}{d_\beta}=\int_{\Upsilon_\text{p}} \pd{\tilde{\mathbb{C}}_{ijkl}}{d_\beta}-\pd{}{d_\beta}
\left(\tilde{\mathbb{C}}_{ijst}J_{nt}^I\pd{\xi^{kl,I}_s}{\bar Y_n}\right)\d {\bar{\mathbf{Y}}}.
\eeq
Note that the microscale design variables controls the cell configuration $\Upsilon_{\text{p}}^{\text{s}}$, and thus the piecewise materials constant $\tilde{\mathbb{C}}$ given by Eq.~\eqref{C_bar}, as well as the third order tensor $\boldsymbol{\xi}^I=\left(\xi_k^{ij,I}\right)$, $i$, $j$, $k=1$, $\cdots$, $n$, solved from Eq.~\eqref{equilibrium_micro_zoning}. By adopting a similar treatment as in the so-called adjoint methods \cite{LiuST_SMO2002}, one only needs to consider the derivative of $\tilde{\mathbb{C}}$, and the sensitivity with respect to the microscale variables can be finally expressed by
\beq \label{sensitivity_p20}
\begin{aligned}
\pd{\mathbb{C}^I_{ijkl}}{d_{\be}} &= \int_{\Upsilon_\text{p}}\left(\delta_{ip}\delta_{jq}-J_{nq}^I\pd{\xi^{ij,I}_p}{\bar Y_n}\right) \times \\
& \qquad \pd{\tilde{\mathbb{C}}_{pqst}}{d_{\be}} \left(\delta_{ks}\delta_{lt}-J_{mt}^I\pd{\xi^{kl,I}_s}{\bar Y_m}\right)\d {\bar{\mathbf{Y}}}.
\end{aligned}
\eeq

\subsection{Numerical implementation of the zoning scheme}
\begin{figure}[!ht]
  \centering
  \includegraphics[width=.48\textwidth]{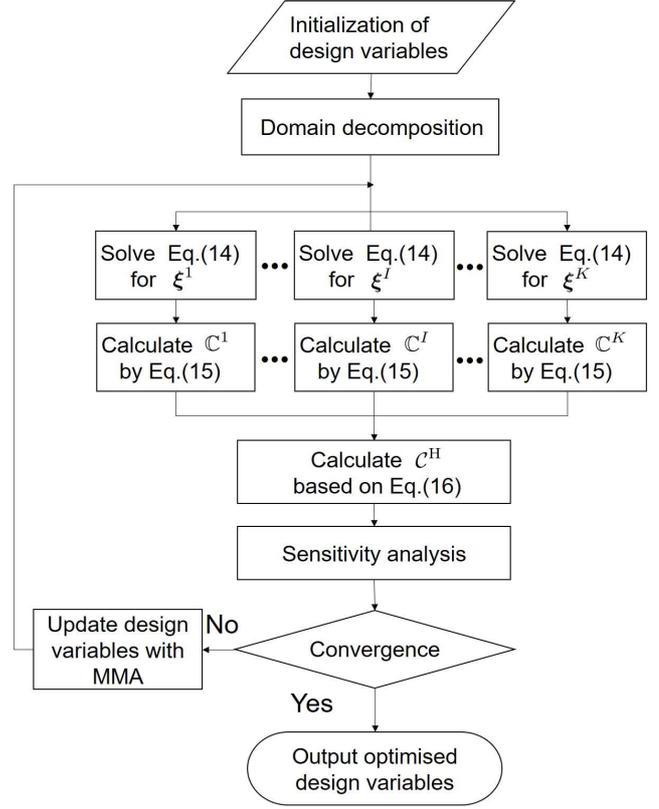}
  \caption{Flowchart of the zoning scheme.}
  \label{flowchart}
\end{figure}

A zoning scheme for optimising the mechanical performance of GMC is thus established, and illustrated in the flowchart~\ref{flowchart}.
An initial guess to the values of $\mathbf{d}$ is firstly given. This is followed by decomposing the overall design domain $\Omega$, with a representative point $\mathbf{x}^I$ assigned to the subdomain $\Omega^I$. With $\mathbf{J}^I=\mathbf{J}(\mathbf{x}^I)$ evaluated, a set of third-order tensor $\xi_k^{ij,I}$ are computed in parallel with reference to Eq.~\eqref{equilibrium_micro_zoning}, and $\mathbb{C}^I$ using Eq.~\eqref{homogenised elasticity_zoning}. Then the macroscopic force balance equation \eqref{equilibrium_homo} is solved, and then the homogenised GMC compliance is approximated by using Eq.~\eqref{Compliance_zoning}. Sensitivity is then conducted referring to Eqs.~\eqref{sensitivity_p1} - \eqref{sensitivity_p20}, and the values of the design variables are updated, unless a convergence criterion is met.

\section{Numerical results and analysis\label{Sec_numerical_examples}}
In this section, numerical results generated with the use of the proposed zoning scheme are presented. After discussion on the general settings for numerical simulations, two-dimensional cases are studied in a systematic way. The 2D numerical examples will be presented in two stages. During the first stage, we mainly examine the performance of the zoning scheme as a way of approximation to the homogenisation formulation (for compliance calculation). It encompasses issues such as the accuracy against the fine-scale results, the choice in the number of subdomains, the efficiency of employing computational parallelism, etc.. During the second stage, we will assess the performance of the zoning scheme as a topology optimiser. This is carried out through a comparison with the benchmark examples obtained by other means \cite{Groen_IJNME2018}. Finally, three-dimensional examples are presented, so as to demonstrate the improved capability of the proposed algorithm in dealing with problems of structural optimisation with a multiple-scale nature.

For all examples examined in this section, linearly elastic response and isotropic material are considered with the parameters chosen as follows: the (nondimensional) Young's modulus $E=1$ and Poisson's ratio $\nu=0.3$. For two-dimensional examples, the specimen's nondimensional thickness in the third dimension is fixed to be 1, and a plane-stress assumption is adopted. Moreover, four-node quadrilateral elements are used for two-dimensional FE simulations, while eight-node hexahedral brick elements are adopted for all three-dimensional simulation examples.

\subsection{Determination of the design variables\label{Sec_num_setting}}
As stated above, design variables on both macro- and microscales should be concerned. But here we keep the microstructural configurations within the fictitious unit cell fixed. The main reasons can be summarised as follows. Firstly, the optimisation with respect to the macroscopic mapping function $\mathbf{y}=\mathbf{y}(\mathbf{x})$ is a major novel point of the proposed method, while the optimisation with respect to the microstructural topology can be treated as a natural extension of classical treatments, such as SIMP \cite{Bendsoe_StructOpt1989,ZhouM_CMAME1991,Sigmund_IJSS1994} and Level-set methods \cite{WangYu_CMAME2003,Allaire_JCP2004}, etc. Secondly, allowing too much degree of freedom for microscale design, as suggested by attempts in a limited number of cases \cite{ZhuYC_JMPS2019}, may lead to a GMC that is too difficult to fabricate. It should also be noted that employing parameterised microscopic cell structures \cite{ChengL_CMAME2019,Groen_CMAME2019,GeoffroyDonders_JCP2020,Groen_arXiv2019} or using TDFs taking multiple level-set values \cite{WangYQ_CMAME2017,ZhangY_CompMaterSci2018,ZhangY_SMO2019} may offer a good trade-off between GMC's variety and manufacturability. At the present stage, however, spatial variation in microscale cell configuration has yet been enabled in the AHTO plus framework. Hence for the purpose of addressing the major issues related to the AHTO plus method, only the macroscopic design variables are in consideration, while the microstructural configurations stay unchanged and are expressed in the so-called MMC-based framework \cite{GuoX_JAM2014}.

A natural way to parameterise the mapping function $\mathbf y(\mathbf x)$ is to employ the form as follows
\beq\label{universe_map}
\mathbf{y}(\mathbf{x})=\sum_{\tau=1}^\kappa h_\tau\Phi_\tau({\mathbf{x}}),
\eeq
where $\Phi_\tau(\mathbf{x})$ comprise of a set of basis functions with $\kappa$ its total number; $h_\tau$ are the corresponding parameters. Hence the mapping function is expressed in a fully explicit form, and the spatial variation is globally controlled by a limited number of parameters $h_{\tau}$. Given these features, a GMC is generated straightforward, soon as all (macroscopic) design variables are evaluated. This is in contrast with de-homogenisation methods where post-processing is needed to numerically resolve a graded microstructural configuration.

In the present work, we adopt polynomial functions of degree-3 as the basic functions, that is,
\beq\label{y_poly}
y_{i}=a_{ij}x_{j}+\frac{1}{2}b_{ijk}x_{j}x_{k}+\frac{1}{3}c_{ijkl}x_{j}x_{k}x_{l}
\eeq
for $i=1$, $\cdots$, $n$. Without loss of generality, symmetry conditions $b_{ijk}=b_{ikj}$, $c_{ijkl}=c_{ijlk}=c_{ilkj}=c_{ikjl}$ are imposed. Thus the Jacobean matrix is expressed in a relatively simple form:
\beq\label{J_poly}
J_{ij}(\mathbf{x})=a_{ij}+b_{ijk}x_{k}+c_{ijkl}x_{k}x_{l}
\eeq
for $i$, $j=1$, $\cdots$, $n$. Incorporating Eq.~\eqref{J_poly} into Eqs.~\eqref{sensitivity_p1a} and \eqref{sensitivity_p1b} gives the sensitivity of the GMC compliance, with respect to the polynomial coefficients $a_{ij}$, $b_{ijk}$ and $c_{ijkl}$ in Eq.~\eqref{y_poly}. Then the algorithm of moving asymptotes (MMA) \cite{Svanberg_IJNME1987} is employed as the optimiser.

Note that the present setup enjoys a design space with a relatively low dimensionality number (18 for two-dimensional cases and 57 for three-dimensional cases). Such a feature, along with the explicitness in the mapping function, helps improve the manipulability of the present AHTO plus approach. Nonetheless, a shortage in the dimensionality number of the design space also brings down the variety of the describable GMCs. This issue will be further discussed in Sec.~\ref{Sec_comparison}.

\subsection{Two-dimensional examples and analysis\label{Sec_result_analysis}}
\subsubsection{A demonstrative example}
\begin{figure}[!ht]
  \centering
  \includegraphics[width=.32\textwidth]{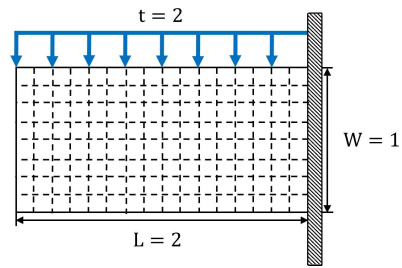}
  \caption{A short beam subjected to uniformly distributed force.}
  \label{short_beam_2D_uniform}
\end{figure}

We start with an example of a short beam subject to a uniformly distributed pressure as shown in Fig.~\ref{short_beam_2D_uniform}.
The right side of the beam is fixed and a uniformly distributed pressure of magnitude 2 is applied on its top. The design domain is of size $2\times1$, and it is divided into $16\times8$ identical square subdomains. The centre of each subdomain is chosen as the corresponding representative points. Here we fix the material distribution in the microscale unit generating cell to be of ``X''-shape, and a $400\times200$ mesh is employed for the computation of the macroscale equation~\eqref{equilibrium_homo}. The volume fraction of the solid material is set to be 30\%.

\begin{figure}[!ht]
  \centering
  \includegraphics[width=.45\textwidth]{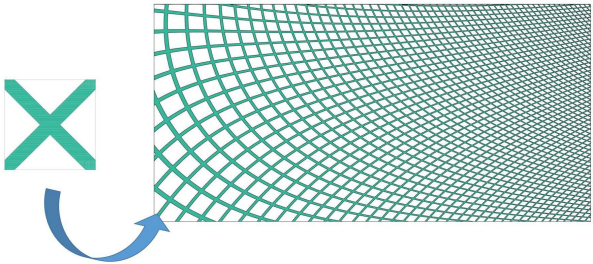}
  \caption{The optimised design obtained with X-shape material distribution in the unit generating cell.}
  \label{uni_X_2D}
\end{figure}

\begin{figure}[!ht]
  \centering
  \includegraphics[width=.45\textwidth]{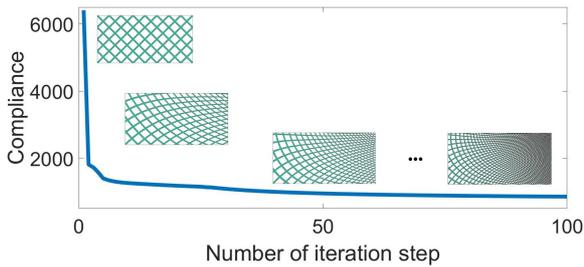}
  \caption{Convergence curve of the GMC compliance.}
  \label{curve_converge_compliance}
\end{figure}

The optimised configuration is shown in the right panel of Fig.~\ref{uni_X_2D}, and the optimisation process is indicated in Fig.~\ref{curve_converge_compliance}. Compared to the (initial) structure composed of spatially periodic microstructure, an optimised microstructure re-arrange their constituting fibres, creating effective conduits transiting the load on top towards the fixed boundary.

\begin{figure*}[!ht]
\centering
\subfigure[ ]{\includegraphics[width=0.45\textwidth]{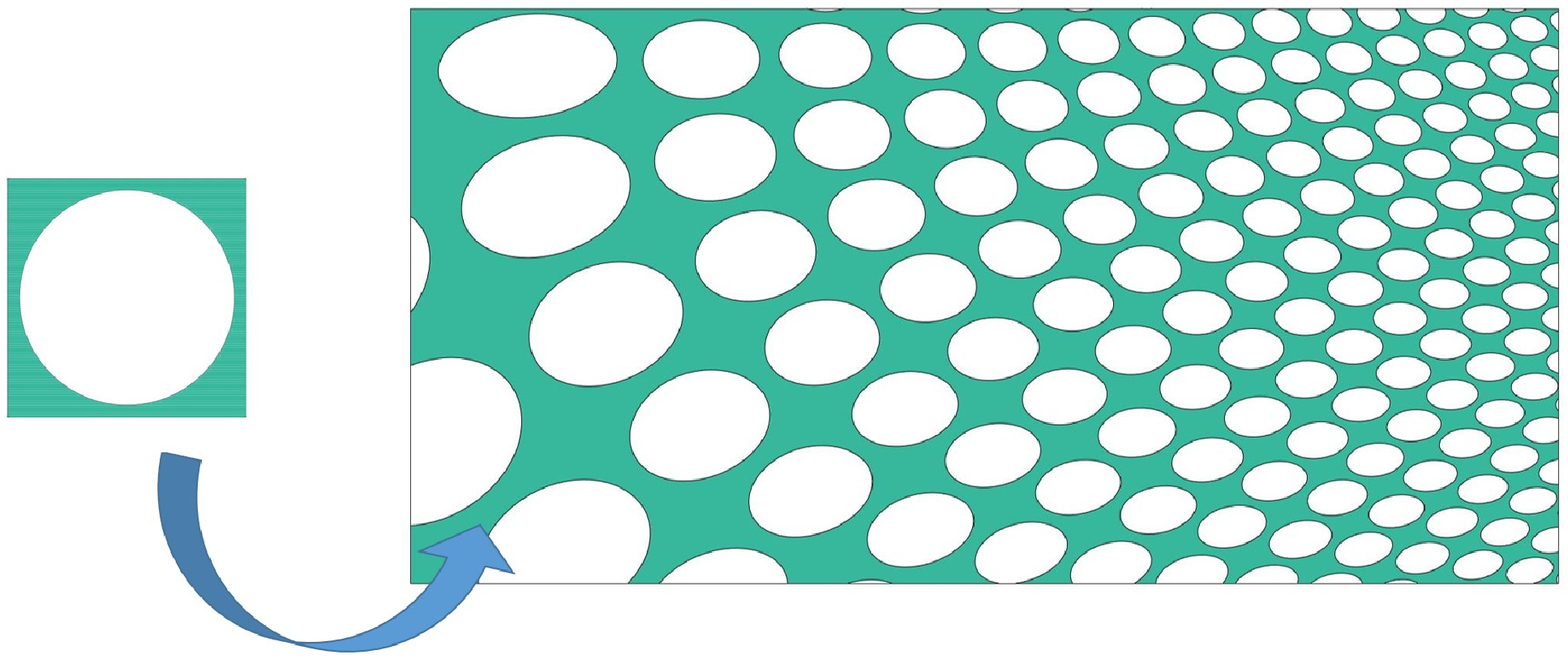}}
\subfigure[ ]{\includegraphics[width=0.45\textwidth]{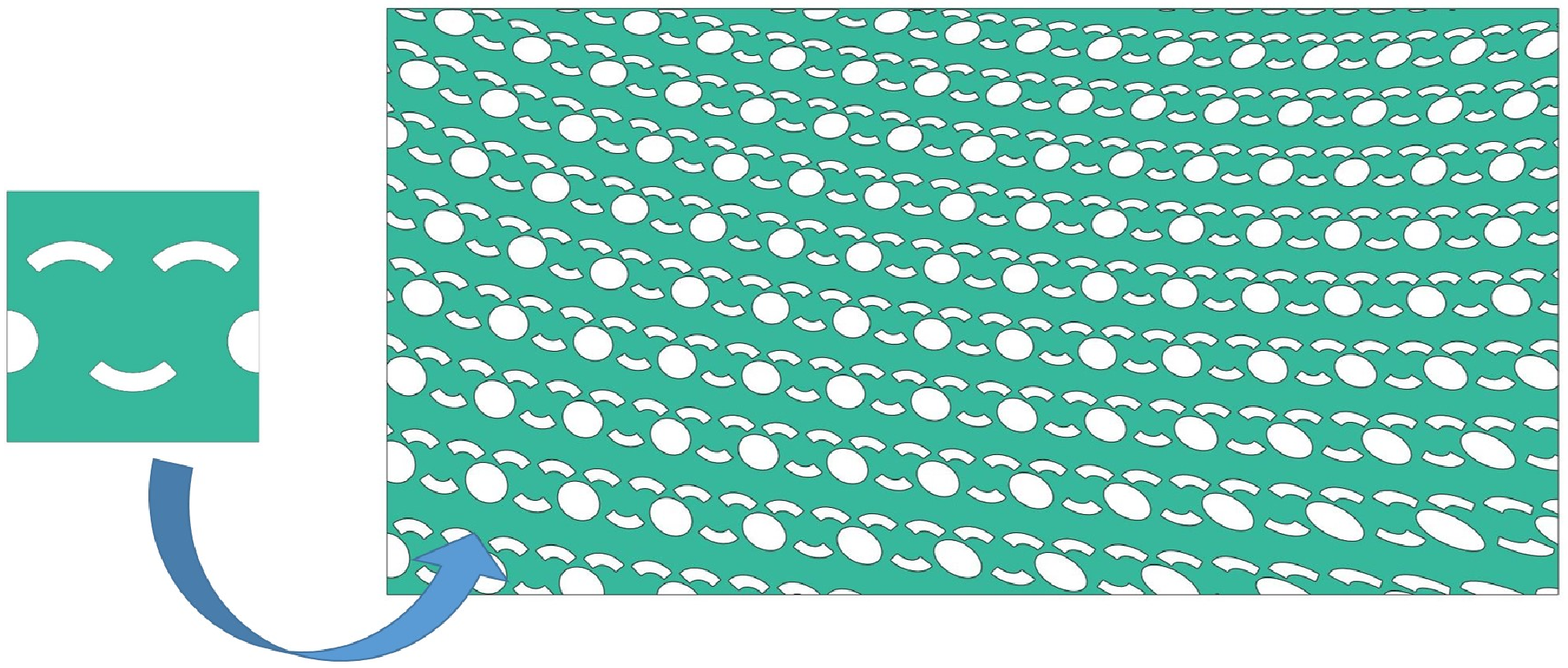}}
\caption{Optimised design with different choices of unit generating cells.}
\label{diversity}
\end{figure*}

With the present homogenisation formulation, the structural compliance is computed to be 852.5, in good accordance with the fine-scale simulation (1600 $\times$ 800) result of 838.6, and the deviation is about 1.6\%. It should be noted that, for direct FE analysis on fine-scale meshes, the top three layers are set to be of full solid, so as to accommodate the uniform load. This only accounts for a volume change no greater than 0.5\%. Further analysis regarding the accuracy of the proposed zoning algorithm against the number of subdomains will be continued in Sec.~\ref{Sec_subdomain_number}.

Note that the present AHTO plus framework does not impose any restrictions over the choice of the unit representative cells. For instance, two other types of unit cell configurations are employed for optimising the loading case given by Fig.~\ref{uni_X_2D}, and the optimised GMCs are shown in Fig.~\ref{diversity}. For the case shown in Fig.~\ref{diversity}(a), the homogenised compliance value (after optimisation) is 510.8, in contrast with the fine-scale result of 491.0 (an error of roughly 4\%). For the case shown in Fig.~\ref{diversity}(b), the homogenised compliance value (after optimisation) is 289.4, in contrast with the fine-scale result of 278.6 (an error of roughly 4\%). From these examples, it is shown that optimisation can still be carried out, even though the directions characterised by the principal stresses to the microstructure are not so obvious. This distinguishes the AHTO plus method from the de-homogenisation approaches, where the building blocks are mainly made of laminate configurations \cite{Groen_IJNME2018,Groen_CMAME2019,Allaire_CompMathAppl2019,GeoffroyDonders_JCP2020,Groen_arXiv2019}. The idea of using laminate lattice structures stems from the discussion that for compliance optimisation of 2D configurations with a single load, the optimal performance should be embodied by rank-2 laminates. This result is still valid for 3D cases involving multiple loads, where optimal results should see rank-3 laminates \cite{Milton_laminate1986,Allaire_Book2002}. Moreover, the use of rank-2 or 3 laminates also facilitates the implementation of the projection method \cite{Pantz_SIAMJCO2008} to finally resolve their GMCs. This is because laminate cells have clear characteristic orientations, and they can be projected straight forward in alignment with the principal stress components within a configuration optimised in a homogenised sense. Nevertheless, as for other loading cases or situations involving multi-physics functions, enabling the choice of more general lattice configuration becomes necessary, and this can be readily achieved in the present AHTO plus framework.

In a number of existing works for GMC optimisation, the TDF associated with the structural topology in a unit cell is preferentially defined in a parametric form, and the spatial variance is achieved fully \cite{ChengL_CMAME2019} or partly \cite{Groen_IJNME2018,GeoffroyDonders_JCP2020,Groen_arXiv2019} by varying the parameter values in space. As for the unit cell shown in Fig.~\ref{diversity}(b), the parameterisation of a ``smiling face'' is not so straight forward. But the face can still be deformed in space within the AHTO plus framework. However, it is also noted that improvement is still necessary for this AHTO plus method, since a change of unit cell configuration in space is not permitted at the present stage, and this points to one direction for its further development.

\subsubsection{Effect by subdomain number\label{Sec_subdomain_number}}
As discussed above, the proposed zoning scheme is expected to deliver a result that is more consistent with the actual situation, if a denser domain partition is employed. However, increasing the number of subdomains poses higher demand on computational resources. In this subsection, we will examine the appropriate partition strategy for the the zoning scheme, i.e., what is the best choice for the number of subdomains?

\begin{figure*}[!ht]
 \centering
  \includegraphics[width=.6\textwidth]{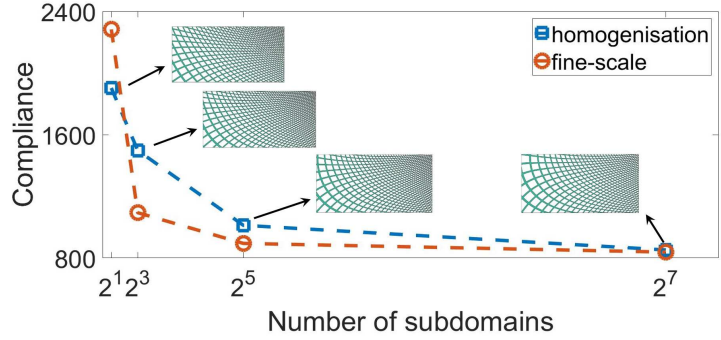}
  \caption{The optimised compliance values (upon homogenisation) against the subdomain numbers are given by squares, and the corresponding optimised GMCs are also indicated. Data represented by circles are the results obtained from corresponding the fine-scale calculations.}
  \label{converge_subdomain_2D}
\end{figure*}

The short beam example with a unit representative cell of ``X''-shape, as depicted in Fig.~\ref{short_beam_2D_unit}, is considered again. Optimisation is carried out with 2, 8, 32 and 128 identical square-shape subdomains, respectively, and the optimisation results (upon homogenisation) are compiled in Fig.~\ref{converge_subdomain_2D}, along with the compliance values obtained from the corresponding fine-scale simulations. Two observation points are addressed here. Firstly, the values of the optimised compliance (upon homogenisation) tend to converge as the number of subdomain increases. Secondly, the zoning scheme generally tends to overestimate the compliance of a GMC, but the deviation becomes negligible when there are 128 subdomains. This demonstrates that the zoning scheme provides an effective mean of approximation to the AHTO plus formulation.

The trends shown in Fig.~\ref{converge_subdomain_2D} also shed more lights on effective implementation of the zoning scheme. Firstly, computational accuracy should be maintained when a subdomain covers no more than 25 finite elements for two-dimensional simulations. Secondly, the algorithm can be further speeded up with the use of a quadtree data structure for two-dimensional cases (and octree for three-dimensional calculations). Initially, optimisation is conducted with an extremely small number of subdomains, until an optimised result is obtained. Then we partition each subdomain by four (for two-dimensional cases), and carry on with optimisation. This processes terminates when the desired subdomain number is reached.

The computational time against the subdomain number is also examined with the data summarised in Table~\ref{tab:1}. The averaged time needed for proceeding one step increases roughly linearly with the subdomain number. We further divide the time needed for carrying out one computational step into three parts, a) computing the homogenised elasticity tensor by solving the microscale equations \eqref{equilibrium_micro_zoning}-\eqref{homogenised elasticity_zoning} with FEA; b) FEA for solving the macroscale equation \eqref{equilibrium_homo}; c) updating the design variables based on the MMA algorithm and the sensitivity results derived in Sec.~\ref{Sec_sensitivity}. The fraction for each part is shown in Table~\ref{tab:1}. Note that in all cases, resolving the microscale problem, that is, part A, takes up most of the computational time in a single step. Moreover, the fraction of the computational time part A is almost 99\%, as the number of subdomains increase to 128. As discussed in Sec~\ref{Sec_introduction_zoning}, the microscale problem governed by Eq.~\eqref{equilibrium_micro_zoning} is inherently suitable for being solved in parallel, and the effect of employing computational parallelism will be examined.
\begin{table*}
\begin{tabular}{ccccc}
\hline\noalign{\smallskip}
No. of & total & microscale & macroscale & data \\
subdomains & time(s) & FEA(\%) & FEA(\%) & updating(\%) \\
\noalign{\smallskip}\hline\noalign{\smallskip}
2 & 13.93 & 52.84 & 17.81 & 29.35 \\
8 & 32.91 & 87.63 & 7.17 & 5.20 \\
32 & 118.95 & 96.42 & 1.98 & 1.61 \\
128 & 459.88 & 98.97 & 0.51 & 0.52 \\
\noalign{\smallskip}\hline
\end{tabular}
\centering
\caption{Solution time decomposition in one iteration step (2D short beam example)}
\label{tab:1}
\end{table*}

\subsubsection{Effect of computational parallelism}
In section~\ref{Sec_theory_parallel}, we have theoretically demonstrated that the proposed zoning scheme should be significantly speeded up with the use of parallel computing treatment. Here we will numerically examine the effect brought by computational parallelism based on the short beam example shown in Fig.~\ref{short_beam_2D_unit}.
\begin{figure}[!ht]
 \centering
  \includegraphics[width=.45\textwidth]{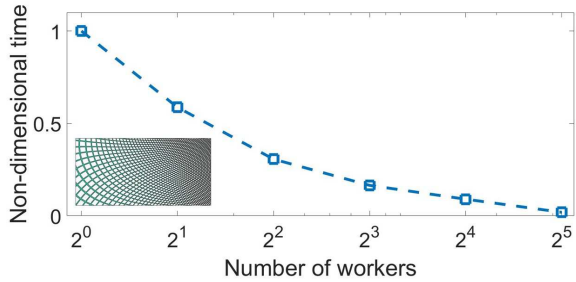}
  \caption{Non-dimensional solution time for each step with different numbers of workers.}
  \label{Time_Kernel_2D}
\end{figure}

We fix the number of subdomains to be 16 and the optimisation is carried out with 1, 2, 4, 8, 16 and 32 workers. The average time needed for proceeding one time step is plotted against the number of workers in Fig.~\ref{Time_Kernel_2D}. Here time is measured in relative terms, i.e. the average time using only 1 worker is set to be 1. It is observed that simply in a MATLAB environment, employing 16 workers saves roughly 80\% of computational time compared to the case of a single worker. This demonstrates the effectiveness of implementing the proposed zoning scheme with the use of parallel computing treatment. The speeding-up effect of using parallel computing is more significant for three-dimensional cases and this will be shown in Sec~\ref{Sec_3D_parallel}.

\subsection{Comparison with existing benchmarked results\label{Sec_comparison}}
\begin{figure}[ht]
  \centering
  \includegraphics[width=.45\textwidth]{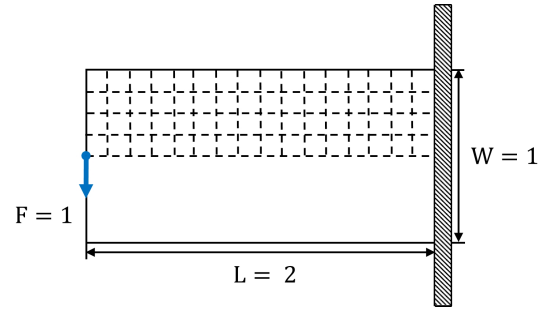}
  \caption{A short beam loaded at the middle point on its left side. The upper-half region chosen as the design domain with symmetric condition imposed across the middle line, and the design domain is analysed and it is divided into 16$\times$4 identical subdomains, each of which is a square.}
  \label{short_beam_2D_unit}
\end{figure}

\begin{figure}[!ht]
  \centering
  \includegraphics[width=.45\textwidth]{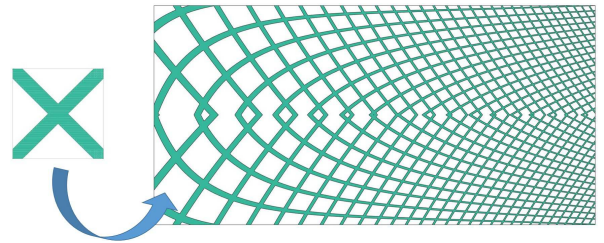}
  \caption{An optimised GMC design for the short beam example. The building block microstructure in the unit generating cell is of X-shape as shown in the left panel.}
  \label{beam_unit_2D_X}
\end{figure}

In this section, we examine the performance of the present AHTO plus method as a tool for structural optimisation of GMC. This time, we consider a short beam fixed on one side as shown in Fig.~\ref{short_beam_2D_unit}. At the middle point on its left edge, a unit vertical load is applied. The macroscopic design domain is a rectangle of length $L=2$ and width $W=1$.
With a symmetry requirement imposed across the middle line, the upper-half region is chosen as the actual design domain, which gets discretised by a $400\times100$ FEs. To implement the proposed zoning scheme, the upper-half domain is divided into 16$\times$4 identical subdomains of square shape, and the centre of each square is chosen as the representative point correspondingly. The building-block microstructure in the representative unit cell for this example, is taken to be of ``X''-shape, as shown in the left panel of Fig.~\ref{beam_unit_2D_X}.

\begin{figure}[!ht]
  \centering
  \includegraphics[width=.45\textwidth]{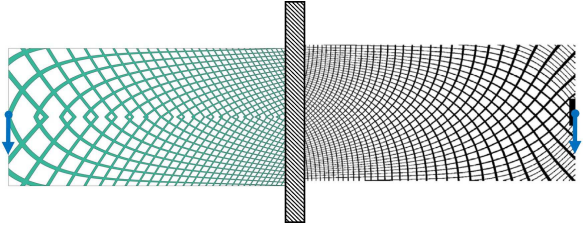}
  \caption{Left: the optimised GMC obtained with the present method; Right: the optimised GMC using the method in \cite{Groen_CMAME2019}.}
  \label{comparison_design_2D}
\end{figure}

Through the use of the zoning algorithm, an optimised GMC is searched for, and the resulting configuration is shown in the right panel of Fig.~\ref{beam_unit_2D_X}. It can be seen that an optimised configuration is composed of fibers connecting the loaded region to the fixed boundaries. The optimised result is also compared with its counterpart obtained using the de-homogenisation method \cite{Groen_CMAME2019}, as shown in Fig.~\ref{comparison_design_2D}.
For the structural compliance, the present AHTO plus formulation outputs a compliance of 282.56, in comparison with the corresponding fine-scale simulation result of 276.38. This is still considerably higher than that ($\mathcal C=228.04$) obtained using the method proposed in \cite{Groen_CMAME2019}. It is worth noting that, for the purpose of making comparison, the microstructural cell is set fixed when using the de-homogenisation approach, but the result can be more optimised if one allows the change in cell structure \cite{Groen_CMAME2019}. This means further improvement is still needed to make the present AHTO plus a more effective TO tool for designing GMCs.

\begin{figure}[!ht]
  \centering
  \includegraphics[width=.36\textwidth]{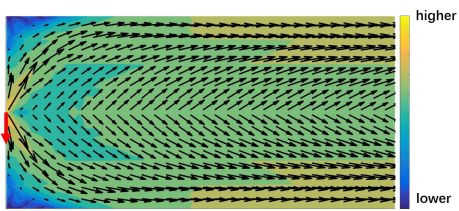}
  \caption{A plot of the energy density distribution (in logarithm term) and the maximum principal stress component (in logarithm term) where the direction is denoted by an arrow.}
  \label{energy_stress}
\end{figure}

Several issues are worth being further discussed, as they provide hints on how to improve the present AHTO plus framework. It can be observed that the optimised GMC in Fig.~\ref{beam_unit_2D_X} does not preserve non-differentiable connection across the middle line, where symmetry condition is imposed. This intuitively explains the reason why it is less optimal compared to its counterpart obtained by the de-homogenisation method \cite{Groen_CMAME2019}. To rationalise the emergence of such non-smoothness, the distribution of elastic energy and the maximum principal stress components are drawn in Fig.~\ref{energy_stress}.
It can be observed that the energy concentration is extremely high near the loaded area. In fact, the contrast in energy density between this region and other places are so high that we have to choose its logarithm values to generate the illustrative contour. It can be read that the materials arrangement is in alignment with the local principal stress at most sites, and a clear path transiting the load to the fixed boundary is also spotted. Moreover, away from the loading point, the middle-line area is far less stressed, and microstructures formed by non-smoothly connected components should still sustain the load. In contrast, the region surrounding the loading point is in a highly stressed state. Thus transition turns smooth towards the loading edge.

Therefore, the GMC given by Fig.~\ref{beam_unit_2D_X} is somehow an optimised result, but is less optimal compared to its counterpart from the de-homogenisation method \cite{Groen_IJNME2018,Allaire_CompMathAppl2019}. The key reason can be found from the expression of $\mathbf{y}(\mathbf{x})$ in Eq.~\eqref{y_poly}. As discussed in Sec.~\ref{Sec_num_setting}, there are only 18 design variables when degree-3 polynomials are chosen to be the basic functions, in contrast to roughly 5000 design variables used in the de-homogenisation approach (for 2D cases). Thus the GMC given by Fig.~\ref{beam_unit_2D_X} should be a sub-optimal solution in a design space of dimension 18, but is still away from the actual optimal point lying in a much higher-dimensional design space.

Moreover, using polynomials to express $\mathbf{y}(\mathbf{x})$ controls the spatial variance of GMC in a global way, while smoothness across the middle line is more like local behaviour. Consequently, when the smoothness is achieved near the loaded area, non-smoothness becomes inevitable in other areas surrounding the middle line.

\begin{figure}[!ht]
  \centering
  \includegraphics[width=.45\textwidth]{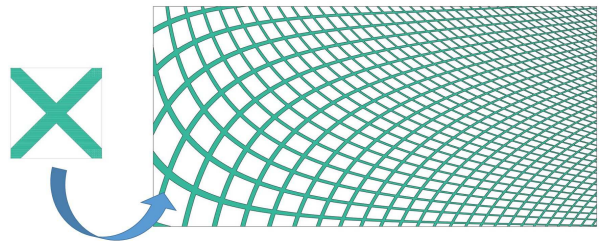}
  \caption{An optimised GMC design for the short beam example. The whole domain is treated as the design domain with no symmetry conditions imposed across its middle line. Smooth connection within the GMC is ensured.}
  \label{full_domain_2D}
\end{figure}

\begin{figure}[!ht]
  \centering
  \includegraphics[width=.45\textwidth]{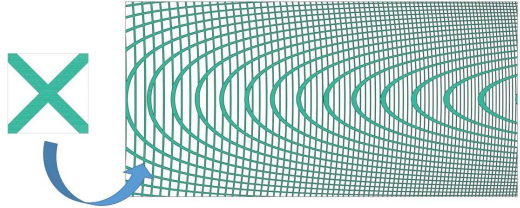}
  \caption{An optimised GMC design for the short beam example. The smooth connectivity across the middle line is ensured by selecting basis polynomials whose normal derivatives across the middle line vanish.}
  \label{smooth_design_2D}
\end{figure}

Here two possible solutions are provided for overcoming the issue over non-smoothness. The first one is not to use the symmetric condition, but to treat the whole domain of size $2\times1$ as the design domain. Fig.~\ref{full_domain_2D} shows the optimised result.
The optimised compliance is 291.15, in comparison with its fine-scale counterpart 282.38. Another possible way to maintain smooth connection across the middle is still using the symmetry assumption, but letting the normal derivatives of $\mathbf{y}(\mathbf{x})$ vanish across the middle line. Correspondingly in Eq.~\eqref{y_poly}, $b_{ijk}=0$, and the optimised configuration is shown in Fig.~\ref{smooth_design_2D}. As the design space contracts further, the resulting compliance rises to 313.04. The examples in Figs.~\ref{full_domain_2D} and \ref{smooth_design_2D} are presented for demonstrating potential means of ensuring smoothness across symmetry boundaries in the AHTO plus formulation, but systematic studies on the parameterisation of the mapping function are still highly in demand.

In summary, further studies should be made along two concurrent directions in order to patch the present AHTO plus method. Firstly, one should manage the trade-off between design freedom and the explicitness of the resulting mapping functions. Secondly, greater attentions should be paid to basis functions posing more weights on local behaviour.

\subsection{Three-dimensional cases\label{Sec_3D_parallel}}
In this subsection, the compliance optimisation of three-dimensional GMC is studied. After several numerical examples are presented, the effectiveness of using parallel computing is then discussed.

\begin{figure}[!ht]
  \centering
  \includegraphics[width=.45\textwidth]{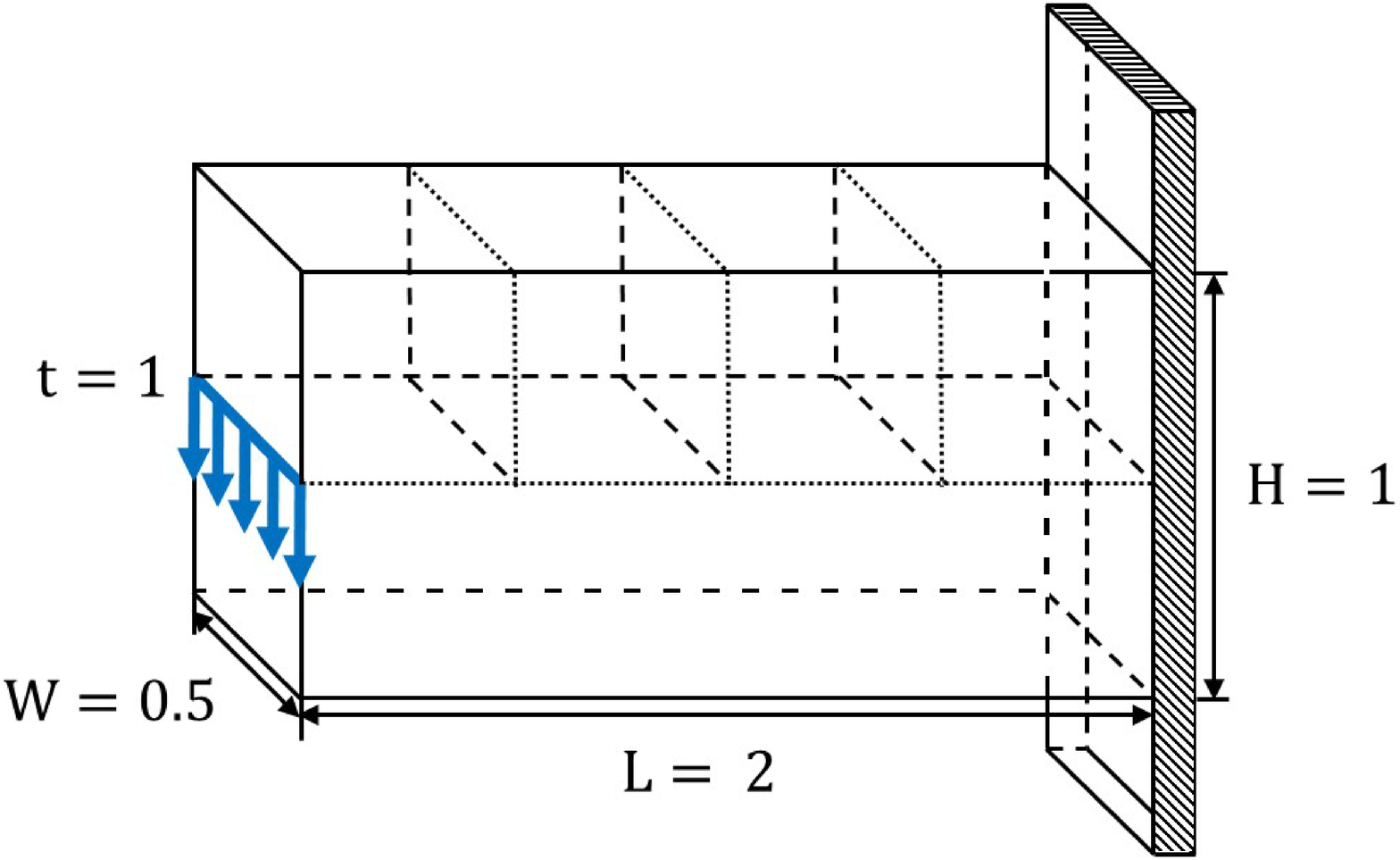}
  \caption{A quasi-two-dimensional short beam example.}
  \label{line_force_3D}
\end{figure}
As shown in Fig.~\ref{line_force_3D}, a short beam of length $L=2$, hight $H=1$ and thickness $W=0.5$ is considered. The beam is fixed on its right side and its left side is loaded with a line force distributed uniformly vertical to its middle plane. Similar to the two-dimensional short beam case, only upper half of this part of the computational domain is considered with symmetry conditions imposed across the middle plane. The homogenised half domain is discretised by a $80\times20\times20$ FE meshes. Limited by the available computational resources, the design domain is divided into (up to) 4 identical cubic subdomains.

\begin{figure}[!ht]
  \centering
  \includegraphics[width=.45\textwidth]{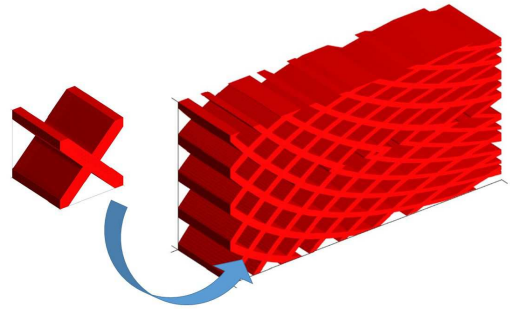}
  \caption{The optimised design obtained with a X-shape generating cell.}
  \label{line_X_3D}
\end{figure}

To validate our 3D algorithm, we first consider a quasi-two-dimensional situation, where the cell distribution is uniform along the direction of thickness. Hence the representative unit cell constitutes two solid components crossing each other to form an ``X''-shape, as shown in the left panel of Fig.~\ref{line_X_3D}.
The optimised GMC is shown in Fig.~\ref{line_X_3D}, and it bears a strong similarity with the corresponding optimised 2D result shown in Fig.~\ref{beam_unit_2D_X}.

\begin{figure}[!ht]
  \centering
  \includegraphics[width=.45\textwidth]{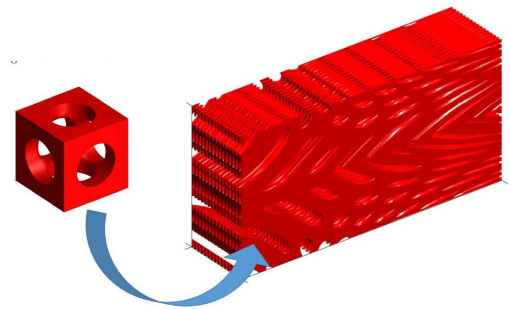}
  \caption{The optimised design obtained with an Orthogonal-shape generating cell and 4 subdomains.}
  \label{line_O_3D}
\end{figure}

Now we consider a more general case, where the unit cell takes a fully 3D configuration, as shown in the left panel of Fig.~\ref{line_O_3D}. Now the problem considered is the same as that shown in Fig.~\ref{line_force_3D}. In the optimised GMC, as shown in Fig.~\ref{line_O_3D}, the unit cells deform themselves in space such that solid materials are distributed along the path connecting the load to the fixed end.

Several critical issues that make 3D cases essentially different form 2D ones will be mentioned. The degree of spatial variance for 3D cases is far greater than the 2D cases. As a result, some of the generated 3D GMCs may be too complicated to be fabricated. Thus in order to ensure structural manufacturability, we suggest that a limited number of basis functions should be adopted when expressing $\mathbf{y}(\mathbf{x})$, the mapping function. For the simulation results presented in Fig.~\ref{line_X_3D}, we exclude the spatial variation along the direction of thickness at the continuum level. Nevertheless, how to effectively choose basis functions for $\mathbf y(\mathbf x)$, as in 2D cases, is still an interesting issue worth being further investigated (elsewhere).

\begin{figure}[!ht]
  \centering
  \includegraphics[width=.45\textwidth]{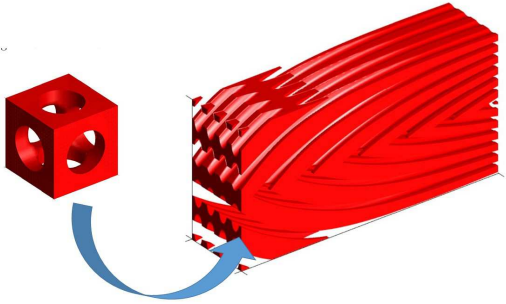}
  \caption{Deformation of the unit cell without imposing any constraints on the determinant of $\mathbf{J}$. In this scenario, the constituting cells may get elongated too much, and it goes beyond the validity range of the AHTO plus formulation.}
  \label{line_Or_3D_none}
\end{figure}

Besides, as 3D optimisation proceeds, the unit cell tends to be elongated to a size that is comparable to the continuum domain size, as shown in Fig.~\ref{line_Or_3D_none}. This makes the result go beyond the validity range of the AHTO plus formulation. To avoid this, one may set bounds to the determinant of $\mathbf J$ defined by Eq.~\eqref{J_poly}. Here we set $1/3<|\det\mathbf{J}|<3$.

Table \ref{tab:2} records the time consumed to carry out one single optimisation step, and the time (in fraction term) needed for processing different stages are also listed. The microscale FEA takes up more than 95\% of the time consumed in one step. This value is even higher than that in the 2D case given in Table~\ref{tab:1}, provided a same number of subdomains. This is because the microscale analysis, governed by Eq.~\eqref{equilibrium_micro_zoning}, calculates for a third-oder tensor $\boldsymbol{\xi}=\left(\xi^{ij}_k\right)$, $i$, $j$, $k=1$, $\cdots$, $n$. When the spatial dimensionality increases, the number of equations involved rises from $2^2$ to $3^2$. Besides, the time (in fraction term) needed for microscale FE analysis increases with the number of subdomains, and the value is more than 99\% with 4 subdomains. Therefore, the use of parallel computing should bring about even higher efficiency for 3D computation.
\begin{table*}
\begin{tabular}{ccccccc}
\hline\noalign{\smallskip}
No. of & total & microscale & macroscale & data \\
subdomains & time(s) & FEA(\%) & FEA(\%) & updating(\%) \\
\noalign{\smallskip}\hline\noalign{\smallskip}
1 & 74.7950 & 96.88 & 2.59 & 0.53 \\
2 & 137.9675 & 98.57 & 1.30 & 0.13 \\
4 & 284.8819 & 99.26 & 0.67 & 0.07 \\
\noalign{\smallskip}\hline
\end{tabular}
\centering
\caption{Solution time decomposition in one iteration step  (3D short beam example)}
\label{tab:2}
\end{table*}

\section{Conclusion and Discussion\label{Sec_conclusion}}
In the present work, a zoning scheme is proposed to facilitate the fast design of configurations filled with graded microstructures concerning their mechanical behaviour. The proposed algorithm effectively improves the applicability of the recently proposed AHTO plus framework, where the representation, stress analysis and topology optimisation of GMCs are systematically integrated. Fueled by the inherent favour of the new asymptotic method for computational parallelism, the simulation efficiency is further improved with the use of the proposed zoning scheme. Moreover, the sensitivity analysis is also carried out so as to equip the AHTO plus method with gradient-based optimisation algorithm. It has been systematically shown with 2D examples that, the mechanical behaviour of a GMC predicted using the proposed algorithm coincides with that from the underlying fine-scale calculations. As a topology optimisater for GMC compliance, the present algorithm enjoys a concise formulation and a small number of design variables, but it outputs an optimised GMC, whose compliance is roughly 20\% higher than its counterpart obtained by using the de-homogenisation methods \cite{Groen_CMAME2019}. Thus effective representation of the mapping function $\mathbf{y}=\mathbf{y}(\mathbf{x})$ is singled out as a key direction for future studies. The considerably high efficiency brought by the use of parallel computing is demonstrated, especially through 3D examples that are not frequently investigated in existing literature.

Equipped with the proposed zoning scheme, the AHTO plus method becomes an effective tool for the description of GMCs and the analysis of their mechanical behaviour. However, as an tool for optimising GMC compliance, it still struggles with hunting GMCs, whose compliances are as optimised as their counterparts obtained by the de-homogenisation methods \cite{Groen_IJNME2018}. Thus it has considerable room for improvement, especially along the following two directions. Firstly, the role played by the basis functions constituting the mapping functions should be examined in depth, and this may give rise to modification over two issues: 1) to ensure smoothness across a boundary imposed with symmetry conditions and 2) to identify a more effective design space where its dimensionality and describability are properly managed. Secondly, the present framework outputs a GMC by deforming a single representative unit cell in space. But representation of GMC from a family of unit cell configurations should also be investigated. For its wider utilisation, the AHTO plus method can be naturally generalised to the optimisation of GMCs for not only mechanical functions, but also other purposes, such as for thermal and acoustics use.

\section*{Replication of results}
All the data underlying the argument in the article are generated by locally devised MATLAB codes consisting of a set of systematically arranged sub-function modules (such as homogenisation etc.). We are willing to satisfy the reasonable and responsible demand for the data and the source codes underpinning the present article.

\begin{acknowledgements}
We thank Ole Sigmund and Jeroen Groen for providing the comparative simulation example shown in Fig.~\ref{comparison_design_2D}. We also thank Ole Sigmund for many valuable comments, without which the article can not reach the present stage. The comments from (anonymous) reviewers during the previous round of review process are also appreciated. The financial supports from National Key Research and Development Plan (2016YFB0201601) from the Ministry of Science and Technology of the People's Republic of China, the National Natural Science Foundation of China (11772076, 11732004, 11821202), the Fundamental Research Funds for the Central Universities (DUT16RC(3)091), Program for Chang-jiang Scholars, Innovative Research Team in University (PCSIRT) are gratefully acknowledged.
\end{acknowledgements}

\section*{Conflict of Interest}
On behalf of all authors, the corresponding author states that there is no conflict of interest.

%
%

\bibliographystyle{spmpsci}      

\bibliography{mybib}

%
%

\end{document}